\documentclass[useAMS,usenatbib,aas_macros]{mn2e}

\usepackage{hhline}
\usepackage{lscape}
\usepackage{graphicx}
\usepackage{colortbl}
\usepackage{amssymb}
\usepackage{amsmath}
\usepackage{natbib}
\usepackage{times}
\usepackage{aas_macros}
\usepackage{threeparttable}

\newcommand{\msun}{~\mathrm{M}_{\odot}}
\newcommand{\zsun}{~\mathrm{Z}_{\odot}}
\def\ltsim{\lower.5ex\hbox{$\; \buildrel < \over \sim \;$}}
\def\gtsim{\lower.5ex\hbox{$\; \buildrel > \over \sim \;$}}

\usepackage{color}
\usepackage{amsmath}
\usepackage{amssymb}

\title[Metal evolution of DCBH hosts]{Metallicity evolution of direct collapse black hole hosts:  CR7 as a case study}
\author[B. Agarwal]{Bhaskar Agarwal$^1$\thanks{E-mail: bhaskar.agarwal@uni-heidelberg.de}, Jarrett L. Johnson$^2$, Sadegh Khochfar$^3$, Eric Pellegrini$^1$, 
\newauthor Claes-Erik Rydberg$^1$, Ralf S. Klessen$^1$, Pascal Oesch$^4$ \\
$^1$Universit{\"a}t Heidelberg, Zentrum f{\"u}r Astronomie, Institut
f{\"u}r Theoretische Astrophysik, Albert-Ueberle-Str. 2, D-69120 Heidelberg \\ 
$^2$X Theoretical Design, Los Alamos National Laboratory, Los Alamos, NM 87545 \\
$^3$Institute for Astronomy, University of Edinburgh, Royal Observatory, Edinburgh, EH9 3HJ\\
$^4$Geneva Observatory, University of Geneva, Ch. des Maillettes 51, CH-1290 Versoix, Switzerland} 

\pubyear{2016}

\begin{document}
\maketitle

\begin{abstract}
In this study we focus on  the $z\sim6.6$ Lyman-$\alpha$   CR7 consisting of clump A that is host to a potential direct collapse black hole (DCBH), and two metal enriched star forming clumps B and C. In contrast to claims that signatures of metals rule out the existence of DCBHs, we show that metal pollution of A from star forming clumps clumps B and C is inevitable, and that A can form a DCBH well before its metallicity exceeds the critical threshold {of $10^{-5}-10^{-6}\ \rm Z_{\odot}$}. Assuming metal mixing happens instantaneously, we derive the metallicity of A based on the star formation history of B and C. We find that treating a final accreting black hole of $10^6-10^7 \msun$ in A for nebular emission already pushes its $H_{160}$ -- [3.6] and [3.6]--[4.5] colours into the 3$\sigma$ limit of observations. Hence, we show that the presence of metals in DCBH hosts is inevitable, and that it is the coevolution of the LW radiation field and metals originating from neighbouring galaxies that governs DCBH formation in a neighbouring {initially} pristine atomic cooling haloes.  

\end{abstract}


\section{Introduction}

Recent observations of the bright Ly$\alpha$ emitter, CR7, at $z\sim 6.6$ \citep[S15 hereafter]{Sobral15a} have sparked a series of studies that aim at exploring its nature. The system is composed of three components: A, B and C, where B and C are thought to host metal-poor (Population II, Pop II) stars, while the composition of A is still under debate. Close to 90\% of the flux can be attributed to component A which also appears to be extremely blue and metal-free. Owing to this, a star cluster of truly metal-free (or Population III, Pop III) stars \citep[see however,][]{YajimaCR7}, or a direct collapse black hole have been proposed as a suitable explanation for  component A (\citealt{Pallotinni15a,DijkstraCR7,SmithCR7}; \citealt[A16 hereafter; ]{AgarwalCR7}\citealt[H16 hereafter; ]{TilmanCR7}\citealt{SmidtCR7,2016ApJ...823..140X,2017arXiv170100814V}). Recently, \citet[B16 hereafter]{Bowler16} have found evidence for metal enrichment in A by interpreting the excess continuum in IRAC Channel 1 (3.6$\mu$m) as an [O\,\textsc{iii}]$\lambda$ 4959, 5007 line. They also reported a 50\% smaller EW for the {\sc{HeII}}$\lambda$1640 line using their new near-infrared data from UltraVISTA.

The aim of this study is to explore if a DCBH in A can form despite the recent presence of metals reported by B16. A recap of the scenario is as follows. 
The formation of a DCBH occurs in pristine atomic cooling haloes exposed to an external Lyman--Werner (LW) radiation field, typically expected to originate from neighbouring galaxies less than $10$ kpc away. The LW radiation is able to dissociate molecular hydrogen, thereby allowing the gas to cool and collapse with atomic H as the main coolant. The resulting isothermal collapse at 8000 K leads to the formation of a $10^{4-5} \msun$ BH. The gas must be nearly metal--free throughout in order to prevent fragmentation into stars \citep{Omukai:2008p113,LatifDust}.

This work builds on our previous model (A16), where we considered A to be polluted (and thus unfit for DCBH) as soon as the first waves of metals from B and C reach it. We now use an analytical model to estimate the actual metallicity evolution of A, given the entire system's mass assembly history. We describe our methodology in Section~\ref{methodology}, followed by the results in Section~\ref{results}. Finally the discussion is presented in Section~\ref{discussion}.

\begin{figure}
\centering
\includegraphics[angle=90,width=0.95\columnwidth,trim={1cm 14cm 1cm 0cm},clip]{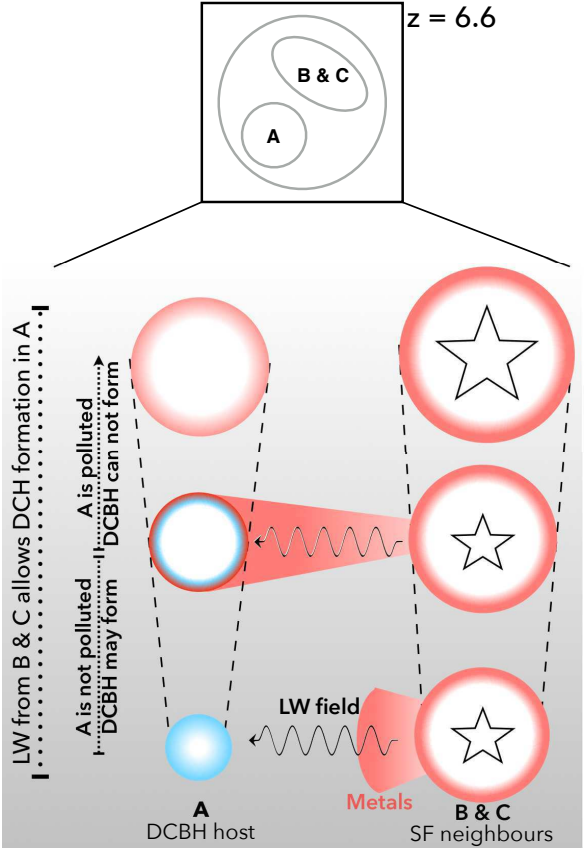}
\caption{Schematic of CR7 and the thesis of the study. The SF components are shown as a single component for the sake of simplicity.}
\end{figure}

\section{Modeling of Metal Enrichment and Observables}
\label{methodology}

The thesis of A16 was that in CR7, the star forming (SF) components B and C produce a LW radiation field that allows for DCBH formation in A. These SF components also produce metals that will pollute A (Fig. 1).
In this study, we use an analytic approach to compute the amount of metals expelled in outflows by B and C, assuming that the outflows travel with a wind velocity, $\rm v_{wind} = 100\ km/s$ \citep[][e.g.]{2015MNRAS.454..238W,2016MNRAS.456.3432G}. Keeping the distance fixed at 5 kpc, we then compute the upper limit on the resultant metallicity of A. This is done for the entire duration that B and C are star forming, which in the model of A16 corresponds to $z\sim 25 - 6$ for an exponentially declining star formation history (SFH) that best fit the observations at $z=6.6$. Additionally, we also consider a constant SFH for B and C in this work that best fits the observations of S15.
The methodology in this study is mostly based on A16 and is summarised below.

\subsection{Mass Assembly History of the System}
We use the same merger history of the system as presented in A16. The dark matter (DM) masses for the individual components are taken as the average of the respective distribution shown in their Fig. 4. Furthermore, we assume that the baryonic mass of the component grows proportional to {the total mass as $\rm f_bM_{tot}$, where $\rm f_b$ is the universal baryon fraction, $\rm M_{tot}$ is the total mass of the halo.}

\subsection{Metal Pollution of A by B and C}

{We concentrate on the outflow and radiation from a single SF component, as the individual nature of B and C is still under debate (S15, B16). B16 report that although the stellar populations in B and C are similar in their observed masses at $z=6.6$ ($10^{9.3}\msun$ vs. $10^{9.7}\msun$), one is considerably younger than the other (200 Myr vs. 500 Myr). The total stellar mass of the SF component considered here is consistent with B16. We also include a case with a constant SFH to better reflect the recent findings of B16, where the age of SF component is $\sim$500 Myr. Thus, henceforth, {the SF component, of SFc,} refers to B and C as a whole and should have no qualitative implications on our findings.}

In order to relate the SFH to the amount of metals produced in the outflow, we first use the parametrisation for the the mass loading factor $\eta$,

\begin{equation}
\rm \eta = \frac{\dot M_{outflow}}{\dot M_\star},
\end{equation}

\noindent {where $\rm \dot M_{outflow}$ is the outflow rate and $\rm \dot{M}_{\star}$ is the star formation rate.} Combining this expression with the results from the {First Billion Years simulations \citep[e.g.][]{Agarwal14, Andrew14, JP2015} that show a correlation between $\rm \eta$ and the stellar mass $\rm M_{\star}$, we obtain (Dalla Vecchia et al. in prep.) }

\begin{equation}
\rm \dot M_{outflow} = 10^{6.3} \, {\dot M_\star} \, \left(\frac{M_{\star}}{\msun}\right)^{-0.8}\frac{\msun}{yr}
\end{equation}

{A general scaling of this form is required to reproduce the low-mass-end of the galaxy stellar mass function and similar scalings have been found in simulations with alternative sub-grid models for feedback \citep[e.g.][]{Keller2016,Muratov2016}.} 

Having obtained the total mass in outflows from B at each redshift, we can compute the mixing time scale motivated by the results of \citet{Cen:2008p841} and \citet{Britton15}. As an upper limit, we assume that metal mixing is efficient and instantaneous in A (for a detailed discussion we refer the reader to the Appendix). {The metals produced in the SF component are spread over a uniform sphere with a radius equal to the separation, D, between A and SFc, i.e 5 kpc. The metallicity of A is obtained from the intersection of this sphere with A's projected circular size, computed using its gravitational binding radius R$^i_{bind,A}$. At any instant $i$, define this dimensionless intersection term as}

\begin{equation}
f^i =\rm  min\left\{0.5, \left(\frac{R_{bind,A}^{\textit i}}{2D}\right)^2\right\},
\end{equation}
 {where $\rm R^\textit{i}_{bind,A}\sim GM_{DM,A}v_{wind}^{-2}$ is the gravitational binding radius of A at timestep $i$. The above formulation allows for up to 50\% of the metals or outflow mass to be captured by A. The time it takes for the metal wind from SFc to reach A is $\rm \Delta_{\mathit{wind}} = {D}/{v_{\mathit{wind}}} = 50\ \rm Myr$. Finally the metallicity of A at any instant, accounting for the wind arrival delay, is computed as}

\begin{equation} 
\rm Z_{A}^{i  + \Delta_{wind}} = \frac{\sum\limits_{t=0}^\textit{i}{\mathit{f} M_{metals,SFc}}}{{M_{baryon,A}^{\textit i} + \sum\limits_{t=0}^{\textit i}\mathit{f} M_{outflow, SFc}}}\ Z_{\odot},
\label{eq.met}
\end{equation}

\noindent where $\rm {M_{metals,B}}$ is the mass produced in metals by SFc at a given time step, $\rm M_{outflow,B}$ is the mass produced in outflows from SFc at a given timestep, $\rm M^\textit{i}_{baryon,A} = f_bM_{tot,A}$, and is the baryonic mass of A at timestep $i$. We estimate the mass in metals by simply multiplying the yield factor, y, with the stellar mass $\rm M_{\star}$. The yield factor depends on the initial mass function (IMF) where y= 0.016 for Salpeter or y=0.032 for Chabrier \citep[e.g.][]{MadauDickinson2014}.

Recent simulations have shown that a DCBH may form in an atomic cooling halo exposed to LW radiation, if its metallicity is below a critical threshold {of Z$_{cr} \sim 10^{-5} - 10^{-6}\ \rm Z_{\odot}$ \citep{Omukai:2008p113,LatifDust}. The lower limit corresponds to a case where both metals and dust are present, whereas the upper limit corresponds to the case of only metals being present. We refer the reader to the Sec. 4 for more details on dust.} The window of opportunity, i.e. the period of time during which DCBH formation in A is possible due to the LW radiation field from SFc is already identified by A16. Thus it remains to be seen if this critical metallicity threshold is exceeded, or not, before the window of opportunity for DCBH formation has elapsed in A.

\section{Results}
\label{results}

\begin{figure*}
\includegraphics[angle=90,width=\columnwidth,trim={2cm 2cm .8cm .55cm},clip]{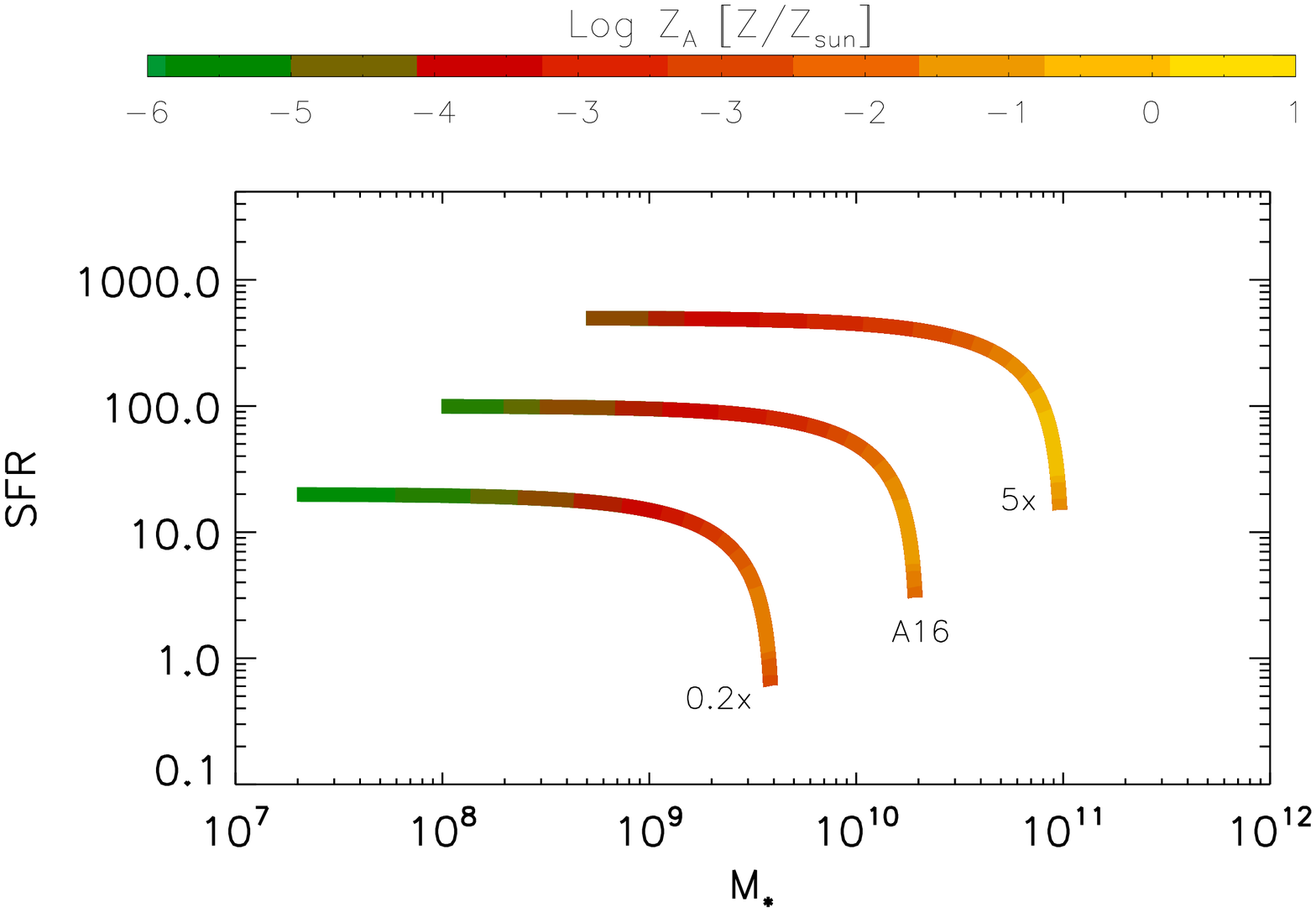}
\includegraphics[angle=90,width=\columnwidth,trim={0.5cm 0.5cm 0.5cm 0.5cm},clip]{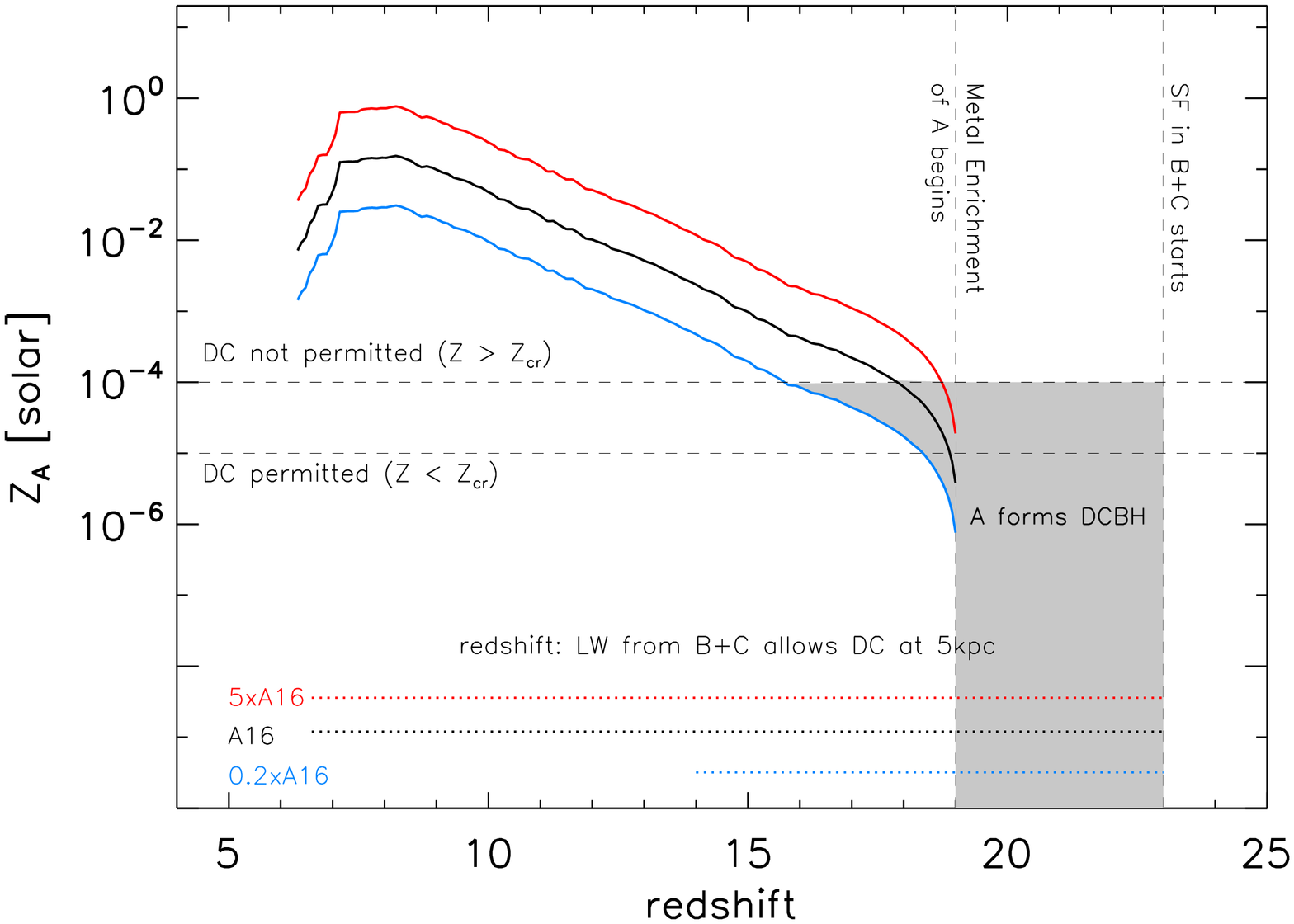}
\caption{The metallicity of clump A, given an exponentially declining SFHs of the SF component in CR7. \textit{Left:} Metallicity of A as a function of the SFH (as presented in A16), scaled up and down by a factor of 5. \textit{Right:} Redshift evolution of metallicity in A and the reduced window of opportunity for DCBH formation, accounting for metal enrichment. {Solid lines are colour coded as per the scaling of the SFH, with black being the fiducial case. The dotted lines of corresponding colour indicate the time window where the LW radiation from the SFc is sufficient for A to form a DCBH. The shaded region in grey bound by the solid lines and the $Z_{cr} = 10^{-4}\zsun$ limit is where A may form a DCBH for the given SFH. For all three cases, A has 50 Myr ($19<z<23$) to form a DCBH before the metals reach it. Even after the metals arrive, A has additional time in all three cases before the metallicity exceeds $Z_{cr} = 10^{-4}\zsun$, as marked by the grey regions that extend to $z<19$.}}
\end{figure*}

\begin{figure*}
\includegraphics[angle=90,width=0.94\columnwidth,trim={2cm 2cm .8cm .55cm},clip]{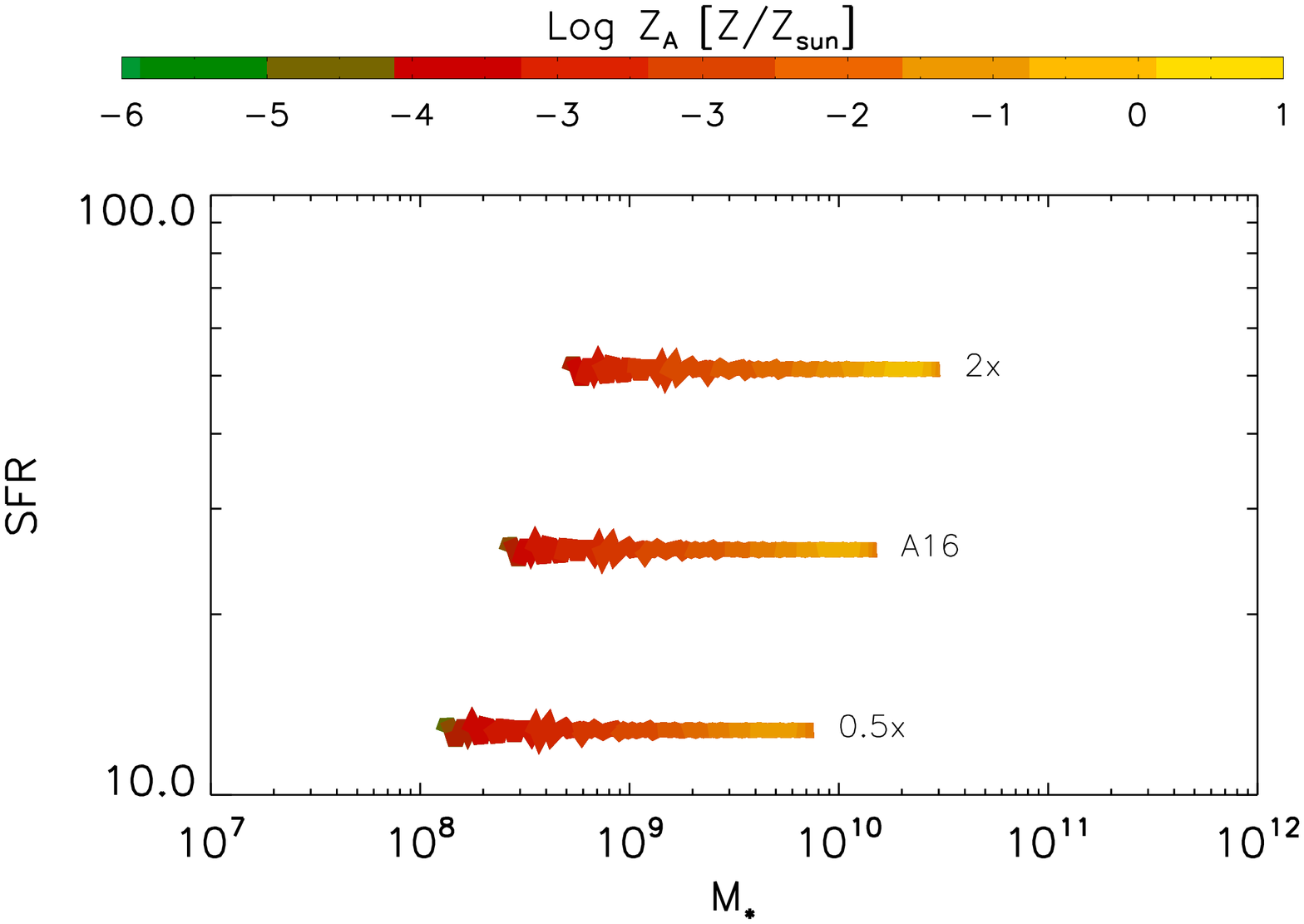}
\includegraphics[angle=90,width=\columnwidth,trim={0.5cm 0.5cm 0.5cm 0.5cm},clip]{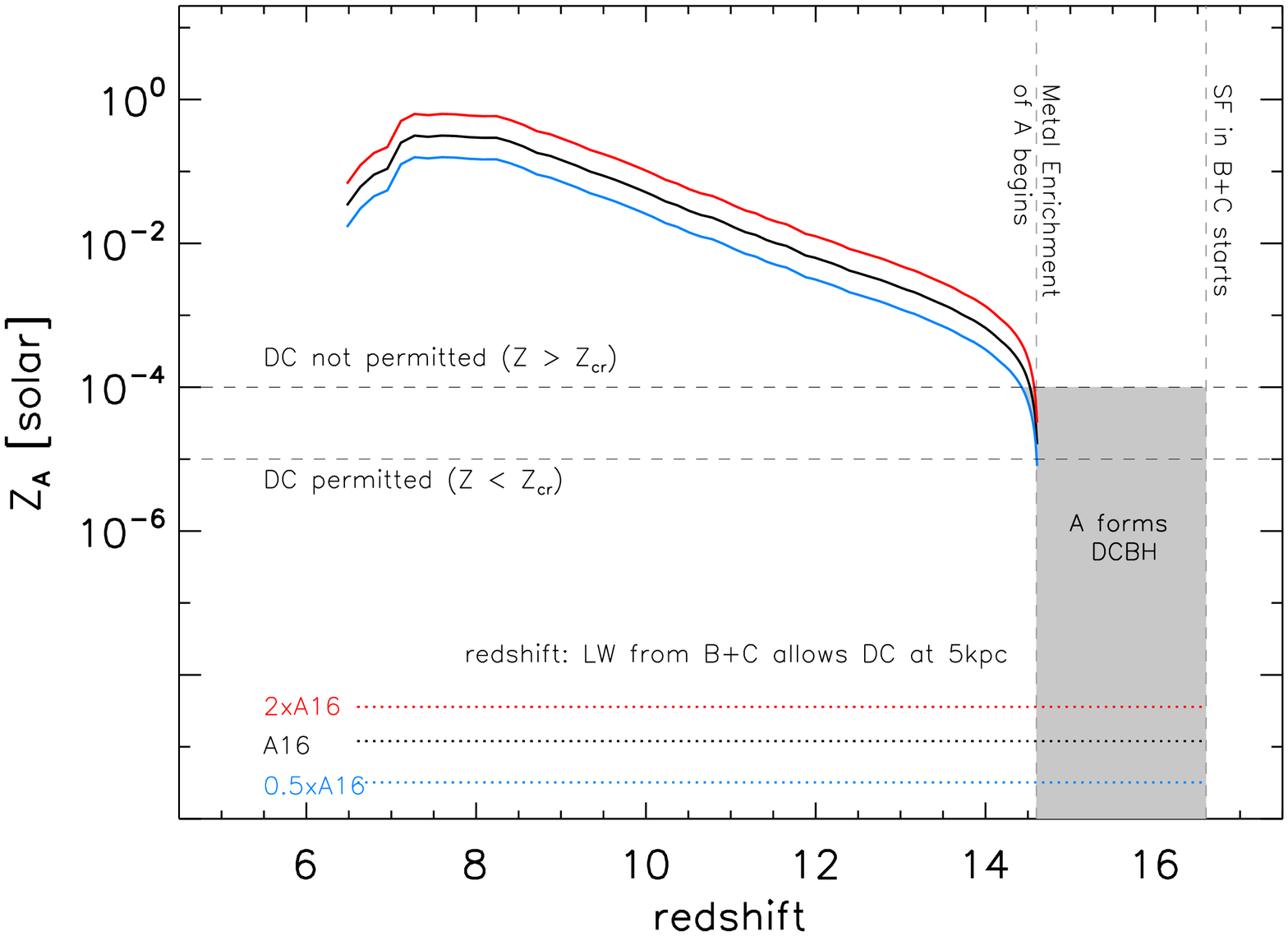}
\caption{ Same as Fig. 2 but for a constant SFH of the SFc in CR7.}
\end{figure*}
\subsection{Implications for CR7}

{Assuming a wind speed of 100 km/s for the outflows and instantaneous metal mixing, we plot the metallicity of A in Fig. 2 and Fig. 3 assuming an exponentially declining, and constant SFH for the SF component of CR7 respectively. We scale the SFH up and down in the SFR-M$_\star$ plane to explore the range of validity of the DCBH scenario. Note that the metallicity in A does not scale linearly with the SFR or M$_{\star}$, as discussed in the previous section. 
The left panels reflect the metallicity of A as a function of the SFR and stellar mass of the SF component. The right panels show the redshift evolution of A's metallicity, taking into account the travel time for winds and the LW window of opportunity.
The right-hand-side panels show that A has ample time, shown as the grey region to form a DCBH before metal enrichment begins, consistent with A16.
In both cases the metallicity to which A is enriched is rather low ( $Z\sim Z_{cr}$), before the window of opportunity for DCBH formation expires- shown in the dotted lines of corresponding colour.  Thus, it is possible that even if the DCBH was to form at the instance of the first event of metal pollution, the gas was not enriched enough to stop DCBH formation. The metal mixing time in A only increases the time window for DCBH formation. Taking the dynamical time of the halo A to be a proxy for metal mixing time scale, at $z=19$ this delay corresponds to an additional $\sim 30\ \rm Myr$, further increasing at lower redshift. Note that at $z=6.6$, all the curves are very close to the theoretical upper limit of metallicity in A derived by H16. Furthermore in the Appendix we show that if the wind speed is instead tied to the escape velocity of halo B, then mixing may always be relative slow and inefficient.} 

{The mass loading factor described in Sec. 2 is best representative of galaxies with M$_{\star}\geq 10^8\msun$. For the same stellar mass (or SFR), a larger mass loading factor would imply a higher outflow rate which would effectively reduce the metallicity described by Eq.~\ref{eq.met}, whereas a smaller mass loading factor might lead to an increase in the metallicity. However, owing to the mass assembly history of A and the separation between the individual components, $f\ll1$, which leads to M$_{\rm baryon, A}\gtsim \sum{f \rm M_{outflow, SFc}}$. Thus the metallicity of A has a stronger dependence on the actual amount of metals produced and the separation between the components, rather than the mass loading factor.}
The conclusions depend on the actual
distance, not the projected distance, between the source galaxy and
the candidate DCBH halo.  In particular, this distance may be greater
than 5 kpc, in which case the metallicity that we estimate for the
host halo would be lower (by the square of the distance).
Thus, we establish here that even in the most effective metal pollution scenario, with $v_{wind} = 100\rm \ km/s$ and $\rm D = 5$ kpc, the DCBH formation epoch derived by A16 remains unaffected, i.e. the DCBH has ample time to form before the critical metallicity in A is exceeded.

\subsection{Reproducing the Observations}

The pollution epoch does not alter the BH's expected growth history. In the case of an exponentially declining SFH in SFc (middle curve Fig. 2), assuming that the BH forms in A at $z\sim19$, the results of A16 remain unaltered. In the case of a constant SFH in SFc (middle curve, left panel of Fig. 3), the DCBH in A can form at $14.5\ltsim z\ltsim16.6$. At the earliest limit, the results remain unchanged. At the lowest possible redshift, i.e. $z\sim 14.5$, a DCBH seed of $10^5 \msun$ allowed to accrete at $25\%$ (45\%) of the Eddington rate can still reach a mass of $10^6 \msun\ (10^7 \msun)$. This still does not change the conlcusions on the estimates of the BH mass derived by A16. Using their estimates, we now derive the observables and check whether they are compatible with the recent work of B16. We consider 4 cases for a BH in A, with a mass of $10^6\rm\ and \ 10^7 \msun$, accreting at 0.5 and 1.0 times the Eddington limit at $z\sim 6.6$. Note that this is the evolved state of the seed DCBH in A, that we assume to have formed at $z>17\ (11.45)$ for the exponentially declining (constant) SFH of B.

Assuming a rest--frame thin--disk incident spectra for the 4 models, we calculate the transmitted and diffuse emission using Cloudy v13.04 \citep{Ferland13}. Following \citet{TilmanCR7}, we calculate 30 realisations over the following unconstrained model parameters: ionisation parameter, metallicities and gas density. Our modeled ionisation parameters range from -2.0 to 0.0. As a default metal abundances we adopt the Orion Nebula {\sc{HII}} stored within Cloudy, scaling all metals and grains from -4.0 to 0.0 in log10. All models are terminated when the gas temperature drops to 5,000~K. The log of the hydrogen number density, $\log_{10} n\rm{(H)}$, was also varied from 1 to 2, but has no significant impact on our results. 

We plot the observable colours (similar to Fig. 2 of B16) of A from our nebular processing in Fig. 4. The input rest--frame spectra are shown as black filled symbols and roughly exhibit the same colour. Already, with the simplest BH spectrum model treated for nebular emission, we are close to the $3\sigma$ limit on the colour of A. Coincidentally, the closest cases all have a metallicity of $0.01\ \rm Z_{\odot}$, consistent with H16, albeit lower BH masses than what was reported in A16. This is interesting as keeping the model--parameters of A16 fixed, and \textit{merely} treating the BH spectral energy distribution (SED) for nebuar emission, our model moves towards the parameter space that is consistent with observations.
{B16 explain the photometric colours of CR7 as a young stellar population with a colour excess in the IRAC 1 band due to strong [O\,\textsc{iii}] line emission.} Our models provide a different interpretation, where the colour is dominated by the "colour" of nebular emission of a fully absorbed AGN incident spectrum spectrum, absorbed and remitted as free-free and bound-free H and He features as well as a contribution by line emission.

The other observed constraint is the modeled EW. Although we use a simpler incident AGN spectrum the predicted colours are well within the observational limits found by \citet{TilmanCR7}, with roughly the same EW vs metallicity and ionization paramter trend. Our assumed thermal emission produces an ionizing spectrum with roughly the same spectral slope. Since the Cloudy models are parametrised with the aid of an ionization parameter, our results are invariant with black hole mass and/or Eddington fraction. 

\begin{figure}
\includegraphics[angle=0, width=\columnwidth,trim={0.2cm 0.2cm 0.2cm 0.2cm},clip]{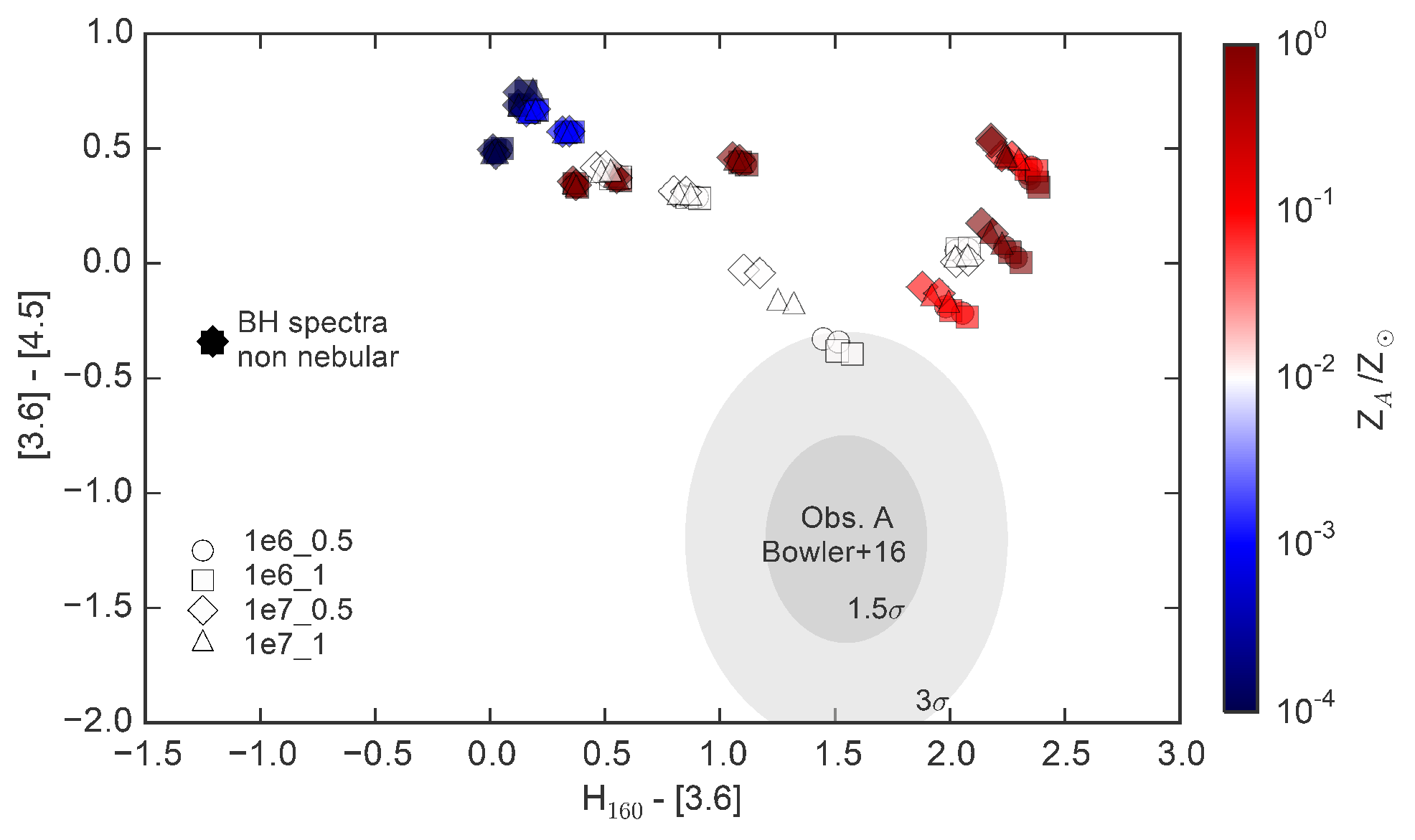}
\caption{The $H_{160}$ -- [3.6] and [3.6]--[4.5] colours of the BH in A, treated for nebular emission. Each symbol style represents a BH mass and the Eddington fraction at which it is accreting (e.g. 1e6\_0.5 stands for a million solar mass BH accreting at 50\% the Eddington fraction). The overlapping black points denote the rest--frame BH SEDs pre--nebular treatment. The observed colours of the BH post nebular treatment are colour coded by their metallicity, where $0.01 \ \rm Z_{\odot}$ (shown in white) corresponds to the theoretical upper limit calculated by H16. The grey regions show the 1.5 and 3$\sigma$ observational limits on A's colour from B16.} 
\end{figure}

\section{Conclusions and Discussion}
\label{discussion}

Recently, B16 report the presence of metals in CR7 which led them to conclude that component A in the system can not be powered by a Pop III or DCBH as both require pristine environments at the time of formation. While the Pop III explanation has been widely refuted \citep{YajimaCR7,Pallotinni15a,TilmanCR7}, we show here that the occurrence of metals in a DCBH host is inevitable. The DCBH host is most likely polluted by the same external galaxies that provide the critical LW radiation field. Using our analytical treatment, we find that although metal rich winds from SF components B and C are able to reach A and pollute it, there is enough time for the DCBH to form in A before its metallicity exceeds $Z_{cr}$. This result is sensitive to the assumption made for the time it takes for metals from B and C to reach A as well as efficiently mix with its gas. We assume throughout our study that metal mixing is efficient and instantaneous. A scenario that does not assume instantaneous metal mixing will only allow for a longer epoch of DCBH formation, thus strengthening our result. 

{The exact nature of dust in the $z>6$ Universe is a far more complicated issue \citep[e.g.][]{Mancini2015,2017arXiv170107869C}, thus we do not explicitly account for pollution by dust. Dust may affect our results in two distinct ways. The first is providing an additional opacity source (beside line opacity) which can lead to efficient radiative driving of the metal enriched outflows due to SF. Studies have shown that radiatively--driven dusty--winds can propagate through the inter--galactic medium with velocities of up to $\sim 1000\ \rm km/s$ \citep{Murray2011}. Second, the presence of dust in CR7 enhances cooling of the gas in A, lowering the critical mass than in the case where no dust and only metals are present. 
The metallicity of the outflow considered in this work can exceed the solar values, and thus the abundance of dust is subject to the details of metal mixing, dust formation and the timescale of outflows. This requires a much more detailed model subject to uncertainties that is beyond the scope of this work. Furthermore, it is unlikely that the dust in the outflow will affect the collapse of the gas in A, as it is likely to be destroyed in the $100-1000 \ \rm km/s$ shocks occurring when the outflows reach A.}

{We explicitly discussed the possibility that component A is powered by a DCBH, however there are other seeding mechanisms as well that can lead to massive seeds, e.g. a rapidly accreting Pop III seed \citep[][Woods et al. in prep., Haemmerle et al. in prep.]{Hosokawa:2013p3513,Alexander14,2016arXiv160904457U}, or a seed resulting from a stellar cluster \citep[e.g.][]{Devecchi_2008,Yajima_2016}. This work is relevant for a scenario where cooling by H$_2$ in a pristine atomic cooling halo needs to be prohibited. The aforementioned alternative scenarios both require star formation at an early stage in A, an unlikely scenario due to the presence of the strong LW field from B and C. Thus it is most likely that the proposed BH in CR7 originated as a DCBH.} 

A DCBH host may also later form or acquire a stellar component, in which case the galaxy might enter a \textit{obese black hole galaxy} (OBG) phase where the accreting DCBH outshines the stellar component \citep{Agarwal13}. These OBGs might be easily distinguishable from the first galaxies by, for e.g. their UV-$\beta$ slope or colours, \citep{Agarwal13,PriyaOBG}. We have not explored the presence of such a stellar population in component A, but do not exclude this possibility. There is large uncertainty surrounding the feedback from the central DCBH on the subsequent  star formation \citep{2014ApJ...797..139A}, thus introducing a range of unknown parameters that would only marginally affect our results in terms of the composition of A.
{Recently, \citet{FabioCR7_2} have presented their model of accretion for the DCBH component in CR7, where the best fit to the observables of B16 is obtained for a M$_{BH} \sim 7\times10^6\msun$ in an $Z=5\times10^{-3} \zsun$ environment. The BH mass is very close to what was predicted by A16, and the metallicity range is consistent with what is presented here, thereby further corroborating the scope of our results.} 
Regardless of the exact nature of the CR7 system it
is important that we understand the range of possible enrichment
histories of galaxies hosting the first accreting black holes, in
order to best be able to identify them and study them with future
missions such as the James Webb Space Telescope.

\section*{Acknowledgements}
The authors are grateful to the anonymous referee for suggestions that considerably improved the scope of this manuscript. The authors would like to thank John Regan, Simon C. O. Glover, Anna Schauer, John Wise, Mark Dijkstra and Ross McClure for useful discussions. BA acknowledges the funding from the European Research Council under the European Community's Seventh Framework Programme (FP7/2007-2013) via the ERC Advanced Grant STARLIGHT (project number 339177).
Financial support for this work was also provided by the Deutsche Forschungsgemeinschaft via SFB 881, "The Milky Way System" (sub-projects B1, B2 and B8) and SPP 1573, "Physics of the Interstellar Medium" (grant number GL 668/2-1).
Work at LANL was done under the auspices of the National Nuclear Security 
Administration of the US Department of Energy at Los Alamos National 
Laboratory under Contract No. DE-AC52-06NA25396.  

\bibliographystyle{mn2e}
\bibliography{babib}

\begin{appendix}

\section{The efficiency of metal mixing}
For simplicity we have assumed that the mixing of the metals ejected from nearby halos occurs immediately and reults in a uniform metallicity in halo A.  However, it is expected that enrichment of the primordial gas in the vicinity of early star-forming galaxies is heterogeneous \citep[e.g.][]{2015MNRAS.451.1190R} and that for the densest gas in cosmological halos, it involves a complex interplay of physical processes \citep[e.g.][]{2016arXiv161000389C,Cen:2008p841,Britton15}.  Here we assess the efficiency of metal mixing by comparing the strength of the self gravity of halo A, which acts to resist mixing with the dense central gas, to the ram pressure of the metal-enriched wind, which acts to entrain the halo gas and enhance mixing.  This ratio is given by Cen \& Riquelme (2008) as the the parameter $\psi$ defined below: 

\begin{equation}
\rm \psi = \left(\frac{1+z}{11} \right)\left(\frac{\rm V_s}{22\ \rm km\ s^{-1}}\right)^{-2}  \left(\frac{M_{\rm H}}{10^6 \, {\rm M_{\odot}}}  \right)^{\frac{2}{3}} \left( \frac{M_{\rm r}}{M_{\rm H}} \right)^{-4.7} \ {,}
\end{equation}
where $z$ is redshift, $V_{\rm s}$ is the speed of the metal-enriched outflow, $M_{\rm r}$ is the mass enclosed within radius $r$ in halo A, and $M_{\rm H}$ is the mass within the virial radius of halo A.  When this ratio is larger than unity it is expected that mixing will occur efficiently on timescales that may be much longer than the dynamical time\footnote{The dynamical time at the virial radius of halo A is $\ga$ 10$^8$ yr at redshifts $\la$ 25 that we consider here.}, yielding the halo relatively stable to mixing with the metal-enriched outflow.  

Figure A1 shows the value of $\psi$, as a function of redshift, for the case of mixing at the virial radius of halo A (i.e. $M_{\rm r}$ = $M_{\rm H}$).  We consider two different wind velocities, to show the sensitivity of $\psi$ to this unknown parameter.  The green curve in the Figure shows $\psi$ for the case of a constant wind velocity of 100 km/s, while the red curve shows $\psi$ for the case of a outflow velocity equal to the escape velocity of halo B.  At almost all times in both cases, the value of $\psi$ easily exceeds unity, implying that metal mixing is inefficient.  It is only in the case of the constant 100 km/s wind at the earliest times ($z$ $\ga$ 18) that metal mixing is predicted to be the most rapid and to occur within a dynamical time.  
From this, we conclude that our assumption of uniform and instantaneous mixing is likely to be too extreme and that our estimate of the metallicity of the gas in halo A should serve as a strong upper limit.
That said, slow metal enrichment of the gas in halo A may well occur over the entire $\sim$ 6 $\times$ 10$^8$ yr between the formation of the candidate DCBH at $z$ $\sim$ 20 and $z$ = 6.6 where CR7 is observed today.      

\begin{figure}
\includegraphics[angle=0,width=3.4in]{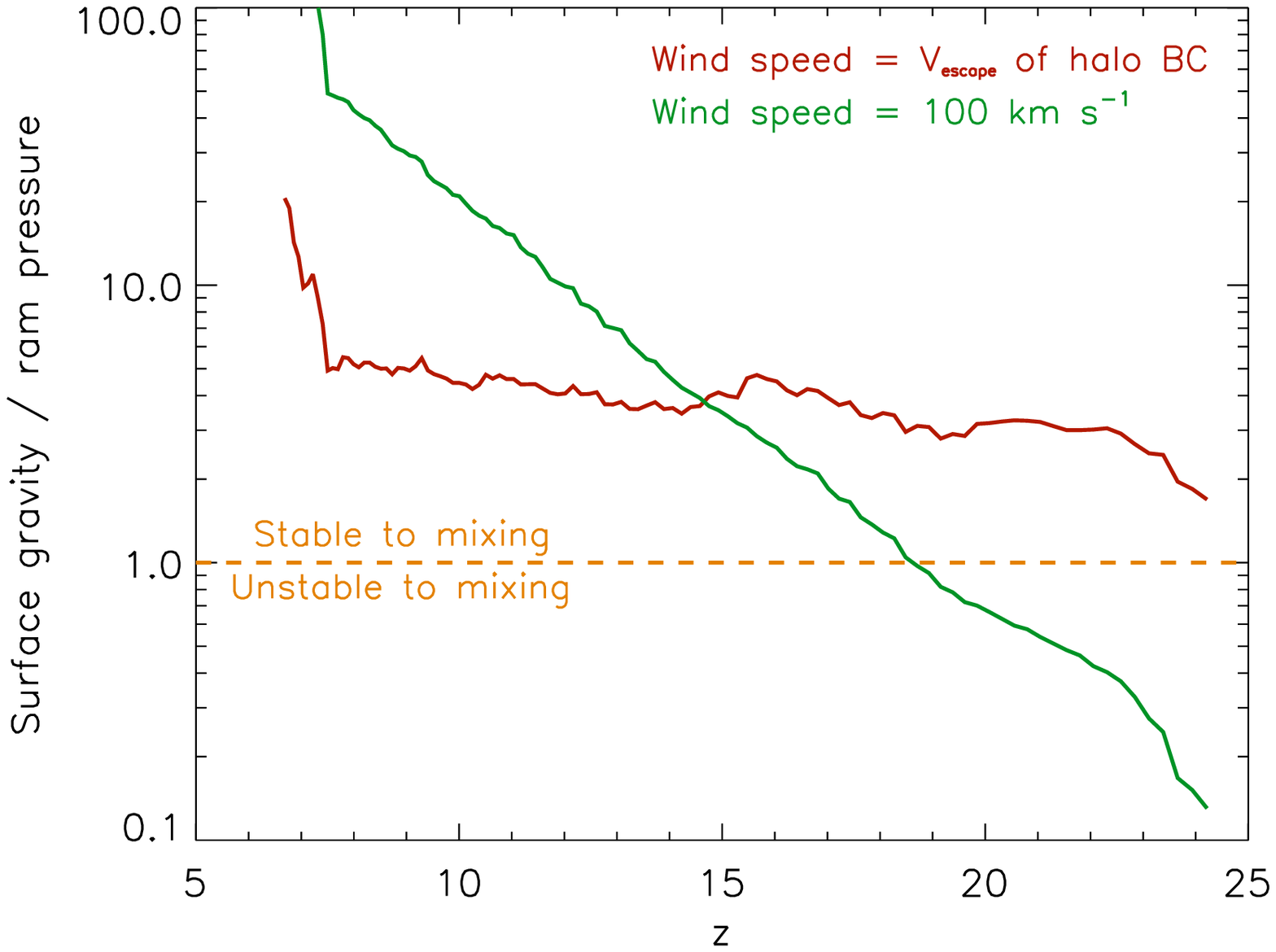}
\caption{The ratio of the force of gravity at the virial radius of halo A to the force of the impinging metal-enriched wind from halo B, as a function of redshift.  }
\end{figure}

\end{appendix}

\end{document}